# Revisiting self-interference in Young's double-slit experiments

Sangbae Kim and Byoung S. Ham*

*School of Electrical Engineering and Computer Science, Gwangju Institute of Science and Technology, 123 Chumdangwagi-ro, Buk-gu, Gwangju 61005, South Korea*

(Submitted on November 18, 2022; bham@gist.ac.kr)

**Abstract:** Quantum superposition is the heart of quantum mechanics as mentioned by Dirac and Feynman. In an interferometric system, single photon self-interference has been intensively studied over the last several decades in both quantum and classical regimes. In Born rule tests, the Sorkin parameter indicates the maximum number of possible quantum superposition allowed to the input photons entering an interferometer, where multi-photon interference fringe is equivalent to that of a classical version by a laser. Here, an attenuated laser light in a quantum regime is investigated for self-interference in a Mach-Zehnder interferometer, and the results are compared with its classical version. The resulting equivalent results support the Born rule tests, where the classical interference originates in the superposition of individual single-photon self-interferences. This understanding sheds light on the fundamental physics of quantum features between bipartite systems.

1.  **Introduction**

Young's double-slit experiments [1] have been studied over the last several decades using not only the wave nature of coherent light but also the particle nature of electrons [2-4], atoms [5,6], and photons [7-10] for quantum mechanical understanding of superposition [11,12]. According to the Copenhagen interpretation, a single photon can be viewed as either a particle or a wave, so that the Young's double-slit interference fringe based on a single photon has been interpreted as self-interference (SI), satisfying the complementarity theory [13]. In other words, the single photon-based interference fringe gives us a quantum mystery simply because of a minimum energy of a photon cannot be split into two parts. Thus, quantum superposition of a single photon in a double slit needs to be interpreted in terms of the wave nature via indistinguishable entities' quantum superposition, otherwise the fringe disappears [7,14,15]. This is the essence of the single photon SI, satisfying the complementarity in quantum mechanics. Without violating generality, a Mach-Zehnder interferometer (MZI) has replaced the Young's double slit due to its experimental feasibilities [7,12,14-20]. The SI of a single photon in an MZI has also been applied for the proof of quantum features such as quantum erasers [7,15], Franson-type nonlocal correlation [16], and N00N state generations with higher-order entangled photon pairs [17-21].

Based on the quantum features of SI in an MZI, indistinguishability between two entities in paired paths has been demonstrated for a polarization basis, where a selection of perpendicularly polarized bases negates the interference fringe due simply to distinguishable entities regardless of their phase coherence [7,14]. So far, those experiments of SI have been performed using quantum particles of entangled photons [12,15,16,18-20]. Keeping in mind that multi-interference among many waves whose frequencies are different results in inherent decoherence, each individual photon's coherence time is supposed to be longer than that of an ensemble. According to the Copenhagen interpretation regarding wave-particle duality or complementarity theory, the particle and wave natures are incompatible or mutually exclusive. Thus, SI should be related with the wave nature of photons, otherwise the physical reality of a single photon in the paired path would be contradictory. In that sense, the first question for SI in an MZI is whether the coherence length of each photon is longer than that of an ensemble of photons. By the way, two-photon intensity correlation, such as the Hong-Ou-Mandel (HOM) effect on a beam splitter [22-26], can be understood as a result of superposition between two individual interferences of single photons [27]. Thus, correct understanding of photon characteristics is important for correct interpretation of the quantum features.

According to the intrinsic bandwidth of photon pairs used for two-photon intensity correlations such as the Franson-type nonlocal correlation [16], N00N state generation [17-21], and HOM effect [22-26], each individual photon's frequency or wavelength is different from the others within the bandwidth, resulting in the ensemble coherence length $l_c$. Thus, it is clear that an individual photon should have a longer coherence length than $l_c$. Here, we experimentally demonstrate that this general understanding of wave optics is not simply true. For this, we use continuous wave (cw) laser-based single photon pairs rather than entangled photons generated from spontaneous parametric down conversion (SPDC) processes [28]. Due to the well-known fact that there is no difference in the interferometric fringe between single photon in a quantum regime and laser light in a classical regime. As a result, we show that the same coherence feature in both single photon and cw cases. As long as a coherent state is defined as a linear superposition of Fock states, a cw coherent state cannot be treated as a quantum entity [29,30].

## 2. Results

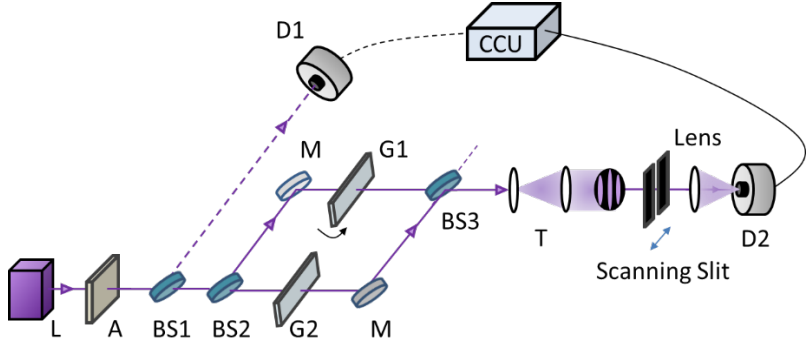

**Fig. 1.** Schematic of quantum superposition for self-interference in an MZI. A: attenuator, BS: beam splitter, CCU: coincidence counting unit, D1/D2: single photon detector, G1/G2: 1-mm thick glass plate, L: 405 nm laser, M: mirror, T: telescope.

Figure 1 shows the schematic of coherent photon-based MZI for the test of SI. For this, we set a reference as a classical bound with a typical MZI fringe using a cw laser at 405 nm of its center wavelendth (Omicron, PhoxX405-120, 200 μW), whose given coherence time and length are $t_c = 0.5\ ps$ and $l_c = 150\ \mu m$, respectively. For a single photon MZI, the 405 nm laser light is attenuated with neutral density filters (OD: 10) until a single photon stream is achieved at a rate of ~2 Mega counts per a second (cps). To confirm photon statistics of the single photon streams, two-photon coincidence measurements are conducted, where doubly bunched photons are split into two parts by the first beam splitter (BS1) in Fig. 1. Because the coincidence detection method eliminates both vacuum states and single photons, only bunched photons are counted for measurements. Bunched photons grouped into three or more are simply neglected, because their generation rates are ~1% compared with the doubly bunched photons. The measured ratio of single photons to the doubly bunched photon pairs ~99% (see Fig. 2).

*2.1. Photon statistics of attenuated light*

Figures 2(a) and (b) respectively show experimental data for both single photons and doubly bunched photon pairs detected by a coincidence counting unit (CCU: Altera, DE2) fed by a pair of single photon counting modules (SPCM: Excelitas SPCM-AQRH 15) of D1 and D2, whose resolving time and dark count rate are 350 ps and < 50 cps, respectively. For Fig. 2, the detector D2 in Fig. 1 is relocated to a position between BS1 and BS2, keeping the same distance as D1 from BS1 (see Fig. S1 of the Supplementary Materials). Each SPCM shows a ~1 Mcps detection rate as shown by the red and black dots in Fig. 2(a), while Fig. 2(b) shows the coincidence detection counting rate at ~12 kcps for the doubly bunched photon pairs. This coincidence counting rate includes three or more bunched photons. Thus, each single photon stream in Fig. 2(a) has ~1% imperfectness of pure single photons for the present experiments of SI. Likewise, Fig. 2(b) also shows ~1 % error due to the Poisson statistics.

To visualize single photon statistics, both detector outputs are fed into a fast (500 MHz) digital oscilloscope (Yokogawa, DL9040). The top panel of Fig. 2(c) shows a pair of single photon streams detected by D1 (yellow) and D2 (green). The middle panel of Fig. 2(c) is an extension of the white box in the top panel. The bottom panels are examples of bunched photon cases detected in the middle panel as shown by the white box (see Figs. S2 and S3 of the Supplementary Information). All detected single photons in the top panel are counted by using a homemade program and compared with those by CCU in Fig. 2(a), resulting in the same number within the error mostly caused by laser intensity fluctuations. The counted number of the bunched photons in Fig. 2(c) is 11 for 1 ms (see Fig. S3 of the Supplementary Materials). Thus, the measured photon characteristics by CCU in Figs. 2(a) and 2(b) is visually confirmed for nearly single photon streams. Considering the 22 ns dead time of the SPCMs, the measured mean photon number in Fig. 2(a) is $\langle n \rangle = 0.04$. Although Fig. 2(a) is Poisson distributed, Fig. 2(b) is sub-Poisson distributed only due to the coincidence detection method eliminating vacuum and single photons. Thus, the self-interference in Fig. 1 belongs to the quantum regime.

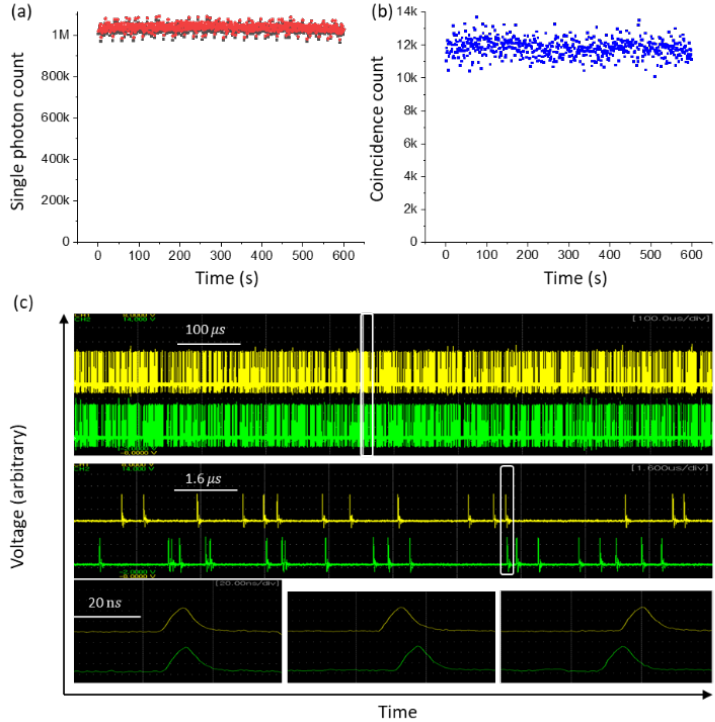

**Fig. 2.** Attenuated photon statistics. (a) Single photons counted by CCU via D1 (red) and D2 (black), where D2 is relocated between BS1 and BS2 in Fig. 1 (see the text). (b) Coincidence detection counts. (c) On the oscilloscope. (Top panel) Corresponding single photon streams for (a). Yellow/green: D1/D2. (Middle panel) Expansion of the white box in the Top panel. (Bottom panel) Expansion of the white box in the Middle panel (see the text).

*2.2. Experimental Design for self-interference*

For the measurements of coherence length of the single photons from the 405 nm laser via coincidence detection, we have designed a slit scanning method for different sets of path-length difference of MZI in Fig. 1. For the path-length control of the MZI, identical glass plates (G1 and G2; 1 mm thickness) are inserted in the MZI paths, where one (G1) is controlled, while the other (G2) is fixed. For G1 control, the rotation angle is adjusted for the variable path-length difference (see Fig. 3(a)). The five different preset path lengths of the MZI correspond to the rotation angle of G1 as shown in Figs. 3(b) and 3(c). The reference position of G1 for $\Delta L = 0$ is confirmed by the maximum of the MZI output fringe. To minimize air turbulence-caused path-length fluctuations, the MZI is covered by a black cotton box, where the fringe stability is remained for several tens of minutes with less than a few per cent fluctuations.

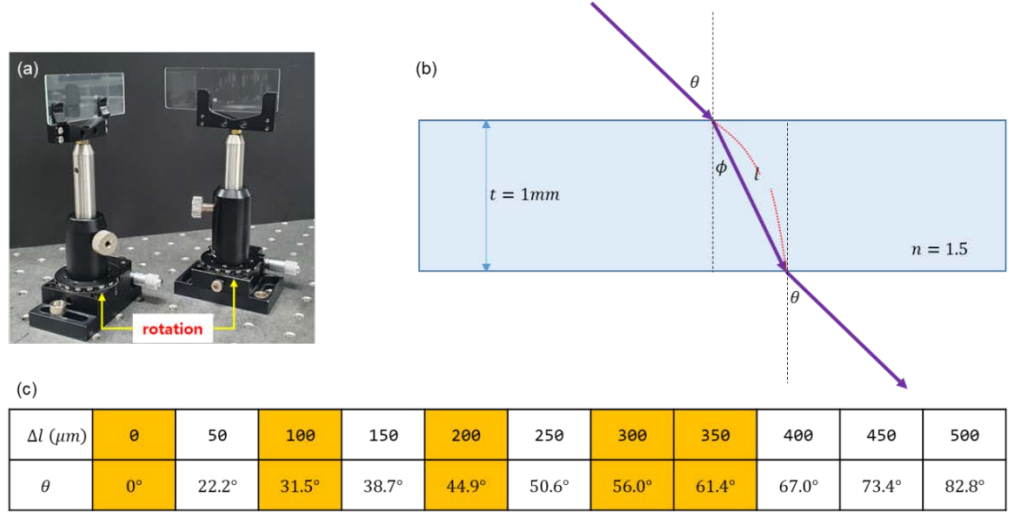

| $\Delta l$ (μm) | 0 | 50 | 100 | 150 | 200 | 250 | 300 | 350 | 400 | 450 | 500 |
|---|---|---|---|---|---|---|---|---|---|---|---|
| $\theta$ | 0° | 22.2° | 31.5° | 38.7° | 44.9° | 50.6° | 56.0° | 61.4° | 67.0° | 73.4° | 82.8° |

**Fig. 3.** MZI path length control. (a) Samples of the glass plate pair inserted in the MZI paths of Fig. 1. (b) A cross section of the glass plate and rotation angle θ with respect to the normal direction. (c) The rotation angle related lengths.

For the MZI fringe measurements, the MZI output light is expanded to ~2 cm in diameter using a telescope T (see Fig. 1). We have designed a homemade single slit of width < 1 mm and inserted it right after the collimated fringe. For the fringe measurements, the slit is repeatedly scanned across the fringe cross-section as shown in Fig. 1. The slit scanning is conducted by a picometer (Thorlabs, Z825B) controlled by a manufacturer supplied software (APT user, Thorlabs) for a fixed range of 20 mm. The spatial resolution of the picometer is 29 nm. The slit scanning time is 2,500 s for single photons and 500 s for cw light. The scanning slit-passed photon (or light) is focused onto the detector D2 by a 10 cm focal-length lens, so that the slit scanning position does not affect the measurement efficiency. The number of MZI fringes is preset to be about six, resulting in each fringe width being much wider than the slit width to neglect potential diffraction effects. Due to 1 s acquisition time by CCU, each scan has 2,500 data points measured. Due to two orders of magnitude bigger number in $\Delta L$ compared with the wavelength λ of the input light, the fringe peaks between repeated scanning trials cannot be coincided.

### 2.3. Self-interference of sub-Poisson distributed single photons

As a reference, conventional MZI output fringes are image captured by a COMS camera (Thorlabs, DC3241M) for the five different ΔLs at the scanning slit position (see the top panels of Fig. 4). The slit scanning is along the dashed line as shown in the first top panel. For the cw laser-based MZI fringes, the single photon detector D1 is blocked, and the SPCM of D2 is replaced by an avalanche silicon photodiode (APD: Thorlabs, APD-110A). This APD is connected to the fast digital oscilloscope to store the measured data. The middle panels show the related cw data. On the contrary, the bottom panels show the corresponding data of single photons in Fig. 2. As analyzed in Fig. 2, the single photon SI measurements have a potential error of ~1% caused by bunched photons.

The acquisition time of each data point for the bottom panels is 1 s as set by CCU under the continuous scanning mode of the slit, resulting in the total data points of 2,500. All data in Fig. 3 are raw without any treatments. For each column of cw and single-photon measurements, the visibility is separately measured from Origin-generated best-fit Gaussian curves (see Figs. S4 and S5 of the Supplementary Materials). Table 1 shows the best-fit Gaussian curve-based visibilities for Fig. 4 (see also Fig. 5).

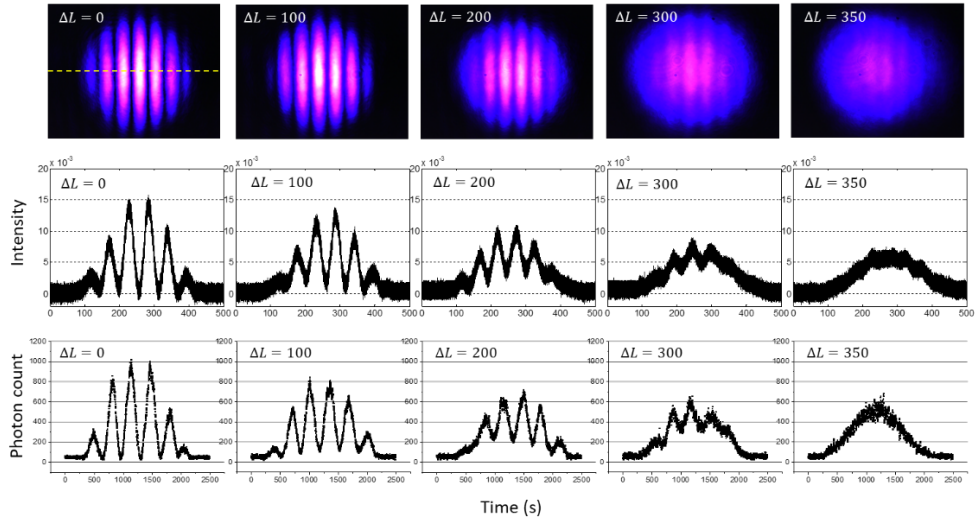

**Fig. 4.** Observations of MZI fringes for different path lengths in Fig. 1. (Top) CMOS captured 2D images for cw coherent light at 200 μW. (Middle) Slit-scanned fringes for the Top. (Bottom) Slit-scanned fringes for attenuated single photons in Fig. 2. The unit of $\Delta L$ is μm. The horizontal dashed line in the upper left panel is an example of the slit scanning direction for the middle and bottom panels.

**Table 1.** Measured visibilities for Fig. 4. $\Delta L$ unit: μm. CW: continuous wave laser light at 200 μW. SP: single photon.

| $\Delta L$ / Source | 0 | 100 | 200 | 300 | 350 |
|---|---|---|---|---|---|
| CW | 1 | 0.801 | 0.539 | 0.3 | 0.153 |
| SP | 1 | 0.8 | 0.525 | 0.326 | 0.171 |

The measured visibilities for Fig. 4 (see Table 1) are plotted in Fig. 5. Figures 5(a) and (b) show examples of how the Gaussian function fitting is conducted for the visibility measurements in Table 1. Figure 5(c) shows best-fit lines for each visibility set in Table 1, where the red line is for the single photon (SP) case, while the black line is for the cw laser. From Fig. 5(c), the measured coherence lengths for both cw laser and the sub-Poisson distributed single photons are 268 μm and 273 μm, respectively, where the cw laser is well satisfied by the factory specification of the 405 nm laser at $> l_c (= 150\ \mu m)$. To our surprise, both visibility decay lines nearly perfectly coincide each other within the standard deviation $\sigma_{CW} = 0.0214$ and $\sigma_{SP} = 0.0212$, demonstrating that both cases result from the same physics, i.e., single photon SI. This result confirms that cw MZI fringe is originated in linear superposition of SIs, resulting in no difference between them as shown in Fig. 5(c). Thus, Figs. 4 and 5 are the direct proofs of single photon SI mentioned by Dirac [11] and Feynman [31], such that a single photon does not interfere with others. This result is quite important to understand the origin of quantum behavior in two-photon correlation such as the HOM effect and Franson-type nonlocal correlation.

One might think that the single photon-based MZI fringe in Fig. 5 could be deteriorated by the laser bandwidth. This is quite correct understanding for the MZI system, because the MZI interference fringes are just superposition of individual SIs. Thus, Fig. 5 demonstrates that cavity optics-based spectral bandwidth of single photons results in the decoherence observed. This fact is quite important with entangled photons from SPDC to view the quantum features achieved from cavity optics or atomic physics, whose observed fringes of HOM effects originates in the same physics [25,26]. In a brief summary, the interference fringe in an MZI is only caused by single photon SI regardless of the photon characteristics. This also validates coherent light as a form of Fock state superposition in an interferometric system such as for Hong-Ou-Mandel dip and Franson-type nonlocal correlation.

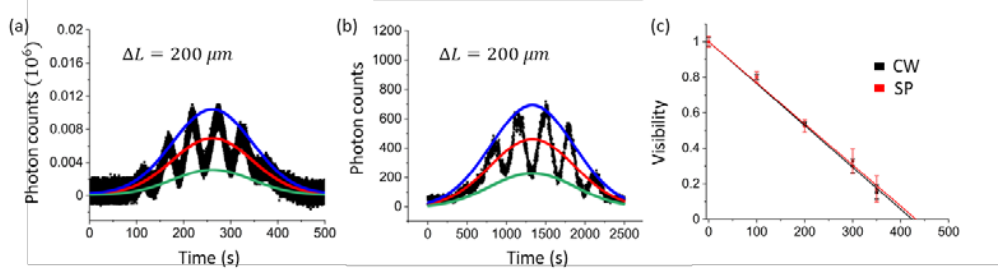

**Fig. 5.** Visibility analysis for Fig. 4. (a) cw and (b) single photons for $\Delta L = 200$ μm. Blue: maximum, Green: minimum, Red: average. (c) Best-fit lines for calculated visibilities in Table 1. Red/black: single photon/coherent light. Standard deviation σ for $\Delta L = 200$ μm: 0.009 (cw), 0.035 (single photon). The best-fit Gaussian curves and error bars are generated using Origin. For details, see Figs. S4 and S5, and Table S2 of the Supplementary Materials. SP: single photon. CW: continuous wave.

*2.4. Data Analysis*

For the visibility measurements in Fig. 5(c) and Table 1, best-fit Gaussian curves are applied, where visibility represents coherence according to the optics theory [32]. For the calculations of coherence length in each case, thus, the visibility decay to $e^{-1}$ is found. For the error bars in Fig. 5(c), actual measured data points are used, where four data points across the highest peak in each panel are averaged for a maximum value, while two sets of four data points from two minima across the highest peak are averaged for a minimum value. For the error bar calculations, the fringe peak does not have to be at center of the Gaussian curve, because each fringe peak is symmetric for the choice of four data points. The four data points correspond to ~1 % variation of the fringe spacing. The increasing errors toward lower visibilities in Fig. 5(c) is due to decoherence-caused Allen deviation [33].

## 3. Discussion

One point worth discussing regarding SI and the Copenhagen interpretation of complementarity theory is as follows. In generalized multi-photon multi-slit experiments for the Sorkin parameter satisfying Born rule in quantum mechanics [34], it is already known that there is no fundamental difference between the classical (coherent states) and quantum interferences [35,36]. In fact, even the observed nonzero Sorkin parameter indicating violation of Born's rule does not necessarily contradict quantum mechanics [37]. As shown in Fig. 5(c), the matched best-fit lines from visibility measurements in Fig. 4 and Table 1 for both coherent light (cw laser) indicating a classical regime and sub-Poisson distributed single photon cases indicating a quantum regime strongly support the fundamental rule of quantum mechanics for quantum superposition in terms of SI. Thus, the SI is the origin of classical interference in an MZI system. Then, the following two questions are raised: 1. Is there any mutual interference among different photons? 2. Does the two-photon interaction such as in the HOM effect result from SI? The answer to the first question must be no according to Fig. 5(c), otherwise there should be accelerated decoherence. The answer to the second question must be yes because of the same role of BS as in the MZI. This is very intriguing and important to correctly understand quantum features.

**4. Conclusion**

Coherent photon-based self-interference (SI) was experimentally demonstrated in an MZI, and the results were compared with that of cw laser light. Regarding the same coherence feature observed in both cases, it was concluded that SI is the origin of the classical interference. This result strongly support our common understanding of quantum mechanics because a coherent state in terms of linear superposition of Fock states should deteriorate the quantum feature of single photons. Thus, our common understanding of the classical interpretation for an MZI fringe is actually based on the quantum feature of the single photon-based self-interference. As claimed by Born-rule tests that no fundamental difference exists between quantum and classical lights, thus, the present observations may shed light on the wave nature-based interpretation of quantum features.

**Funding.** BSH also acknowledged that this work was supported by the ICT R&D program of MSIT/IITP (2022-2021-0-01810), development of elemental technologies for ultrasecure quantum internet and GIST GRI 2022.

**Data availability.** Data underlying the results presented in this paper are available upon reasonable request.

**Disclosures.** The authors declare no conflicts of interest.

**Supplementary Material.** See Supplementary Material for supporting content.